\documentclass[aps,twocolumn,pra,superscriptaddress,showpacs,tightenlines]{revtex4}
\usepackage{epsfig,graphicx,times}
\usepackage{amstext}
\usepackage{amsmath}
\usepackage{amssymb}
\usepackage{graphicx}
\usepackage{latexsym}
\usepackage{bm}

\begin{document}

\title{Non-classical light emission by a superconducting artificial atom
with broken symmetry}
\author{C. P. Sun}
\affiliation{Frontier Research System, The Institute of Physical
and Chemical Research (RIKEN), Wako-shi, Saitama 351-0198, Japan}
\affiliation{Institute of Theoretical Physics, The Chinese Academy
of Sciences, Beijing, 100080, China}
\author{ Yu-xi Liu}
\affiliation{Frontier Research System, The Institute of Physical
and Chemical Research (RIKEN), Wako-shi, Saitama 351-0198, Japan}
\author{L. F. Wei}
\affiliation{Frontier Research System, The Institute of Physical
and Chemical Research (RIKEN), Wako-shi, Saitama 351-0198, Japan}
\affiliation{Institute of Quantum Optics and Quantum Information,
Department of Physics, Shanghai Jiaotong University, Shanghai
200030, P.R. China }
\author{Franco Nori}
\affiliation{Frontier Research System, The Institute of Physical
and Chemical Research (RIKEN), Wako-shi, Saitama 351-0198, Japan}
\affiliation{Center for Theoretical Physics, Physics Department,
Center for the Study of Complex Systems, The University of
Michigan, Ann Arbor, Michigan 48109-1040}

\date{\today }

\begin{abstract}
We propose a novel method to generate non-classical states of a
single-mode microwave field, and to produce macroscopic cat states
by virtue of a three-level system with $\Delta$-shaped (or cyclic)
transitions. This exotic system can be implemented by a
superconducting quantum circuit with a broken symmetry in its
effective potential. Using the cyclic population transfer,
controllable single-mode photon states can be created in the third
transition when two classical fields are applied to induce the
other two transitions. This is because, for large detuning, two
classical fields are equivalent to an effective external force,
which derives the quantized single mode. Our approach is valid not
only for superconducting quantum circuits but also for any
three-level quantum system with $\Delta$-shaped transitions.
\end{abstract}


\maketitle \pagenumbering{arabic}

\section{Introduction}

The symmetry of a quantum system determines the selection rules of
its transitions. For instance, all states of a generic atom must
have a well-defined parity, and one-photon absorption (emission)
due to the electric-dipole interaction can only happen for
non-degenerate states with opposite parities. For second-order
processes, a two-photon transition requires that these states have
the same parities. Thus single- and two-photon transitions between
two given energy levels cannot coexist.

Most investigations so far have focused on either $\Lambda$, or
$\Xi$ (ladder), or $V$-type transitions~\cite{scullybook,jpm} when
studying three-level atomic systems. These notations, defined
according to the transition configuration, are well known to
physicists studying atoms and optics. For example, a $\Lambda
$-type transition atom means that there are optical transitions
from the top energy level to the two lower energy levels,
respectively; however, the optical transition between the two
lower energy levels is forbidden.

The $\Delta$-type three-level systems with cyclic transitions
(CT)~\cite{cyclic}, in which one-photon and two photon processes
coexist, are less common. It is of interest to explore the
possibility of the coexistence of single photon and two photon
processes. For chiral and other broken-symmetry systems, the lack
of inversion center allows the CT to occur in realistic physical
processes~\cite{cyclic}. It has been shown that such quantum
systems can be experimentally implemented by left- and
right-handed chiral molecules~\cite{cyclic}. With CT, the
populations of the different energy levels can be selectively
transferred by controlling classical fields.

In an atomic system, $\Delta$-type transitions can also be formed
~\cite{fleischhauer} by applying three classical pulses: a pair of
Raman pulses and an additional detuning pulse. It was
shown~\cite{fleischhauer} that the physical mechanism of the
cyclic stimulated Raman adiabatic passage is not an adiabatic
rotation of the dark state, but the rotation of a higher-order
trapping state in a generalized adiabatic basis.

Most recently, the microwave control of the quantum states has
been investigated for ``artificial atoms" made of superconducting
three-junction flux qubit circuits~\cite{liu}, which possess
discrete energy levels. The optical selection rule of
microwave-assisted transitions was analyzed~\cite{liu} for this
artificial atom. It was shown~\cite{liu} that the microwave
assisted transitions can appear for any two different states when
the bias magnetic flux is near the optimal point but not equal to
$1/2$ (the value of the optimal point is $1/2$). This is because
the center of inversion symmetry of the potential energy of the
artificial atom is broken when the bias is not equal to $1/2$.
Then, so-called $\Delta $-type or cyclic transition can be formed
for the lowest three energy levels.

The $\Delta$-type transitions can also be obtained from the model
of the single-junction flux qubit~\cite{zhou,zafiris,migliore}. In
any $\Delta-$type artificial atom, the population can be
cyclically transferred by adiabatically controlling both the
amplitudes and phases of the applied microwave pulses. However,
the population transfer in the $\Lambda-$type artificial
atom~\cite{murali} requires that two classical fields induce the
transitions from the top energy level to other two lower energy
levels, and transitions between two lower energy levels should be
forbidden. This condition can be easily satisfied in the usual
atoms due to the electric-dipole transition rule and its well
defined party. However, in artificial atoms, these two fields can
also induce a transition between two lower energy levels when we
study a $\Lambda$-type artificial atom~\cite{liu}. If some phase
conditions are satisfied, $\Delta$-type transitions can be formed
even with only two classical fields. This is a basic difference
between the usual atom~\cite{fleischhauer} and the artificial
atom~\cite{liu}.

Here, we investigate new phenomena of a cyclic artificial atom,
coupled to a quantized microwave field and controlled by two
classical fields. We will explore the CT mechanism to create a
single-mode photon state, or a {\it macroscopic} Schr\"{o}edinger
cat state which is the entangled state between a macroscopic
quantum two level system (macroscopic qubit) and non-classical
photon states. Our approach is robust because the working space is
spanned by the ground state, or the two lowest energy levels, of
the artificial atom. Because the ground state is not easy to be
excited by the environment in low temperature limit. Also our
scheme is more controllable than either $\Lambda$, or $\Xi$, or
$V$-type atoms, since the extra coupling between the external
field and the two lowest energy levels offers a new controllable
parameter.

Our paper is organized as follows. In Sec.~II, we describe how to
model the superconducting flux qubit circuit as a three-level
artificial system with $\Delta -$type (or cyclic) transitions,
which are induced by the microwave electromagnetic fields. In
Sec.~III, we consider the case with large detuning. In this case,
the top energy level can be adiabatically removed and an
effectively driving field can be applied to the single-mode
quantized field, then nonclassical states can be generated by the
driving quantized field. In Sec.~IV, it is demonstrated that the
standard Schr\"odinger cat state, which is an entangled state
between the inner states of the artificial atom and the
quasi-classical photon state, can be generated. Finally, in
Sec.~V, we give conclusions and discuss possible applications.

\section{Model and Hamiltonian}

\begin{figure}
\includegraphics[bb=80 406 426 691, width=4.4cm,clip]{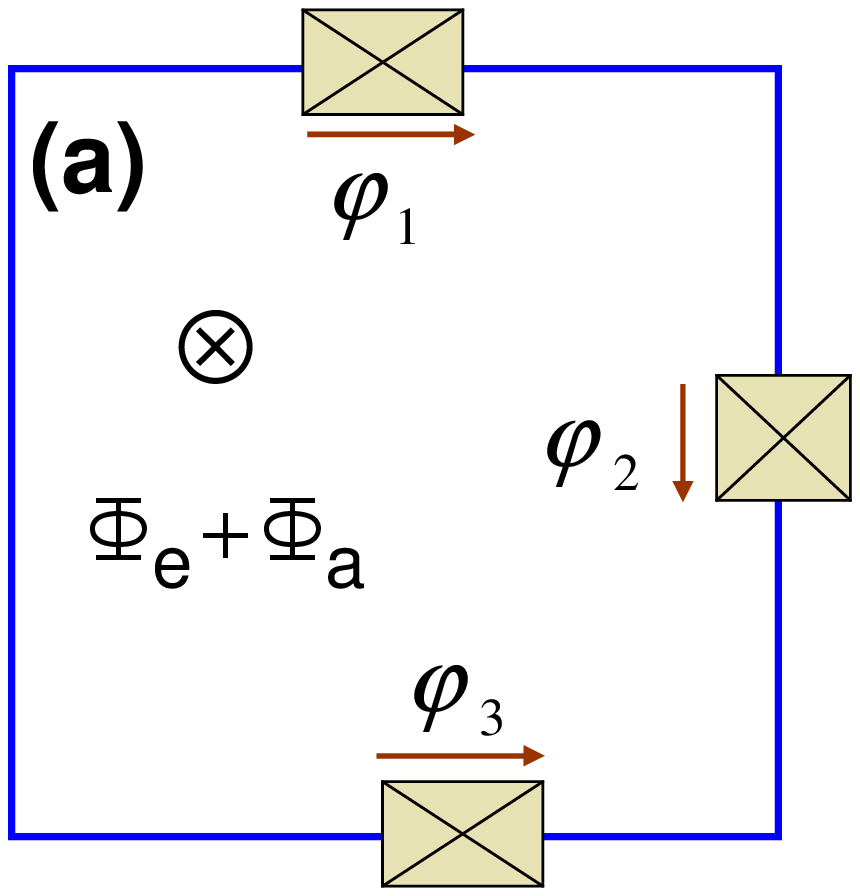}
\includegraphics[bb=40 183 541 597, width=4.1cm, clip]{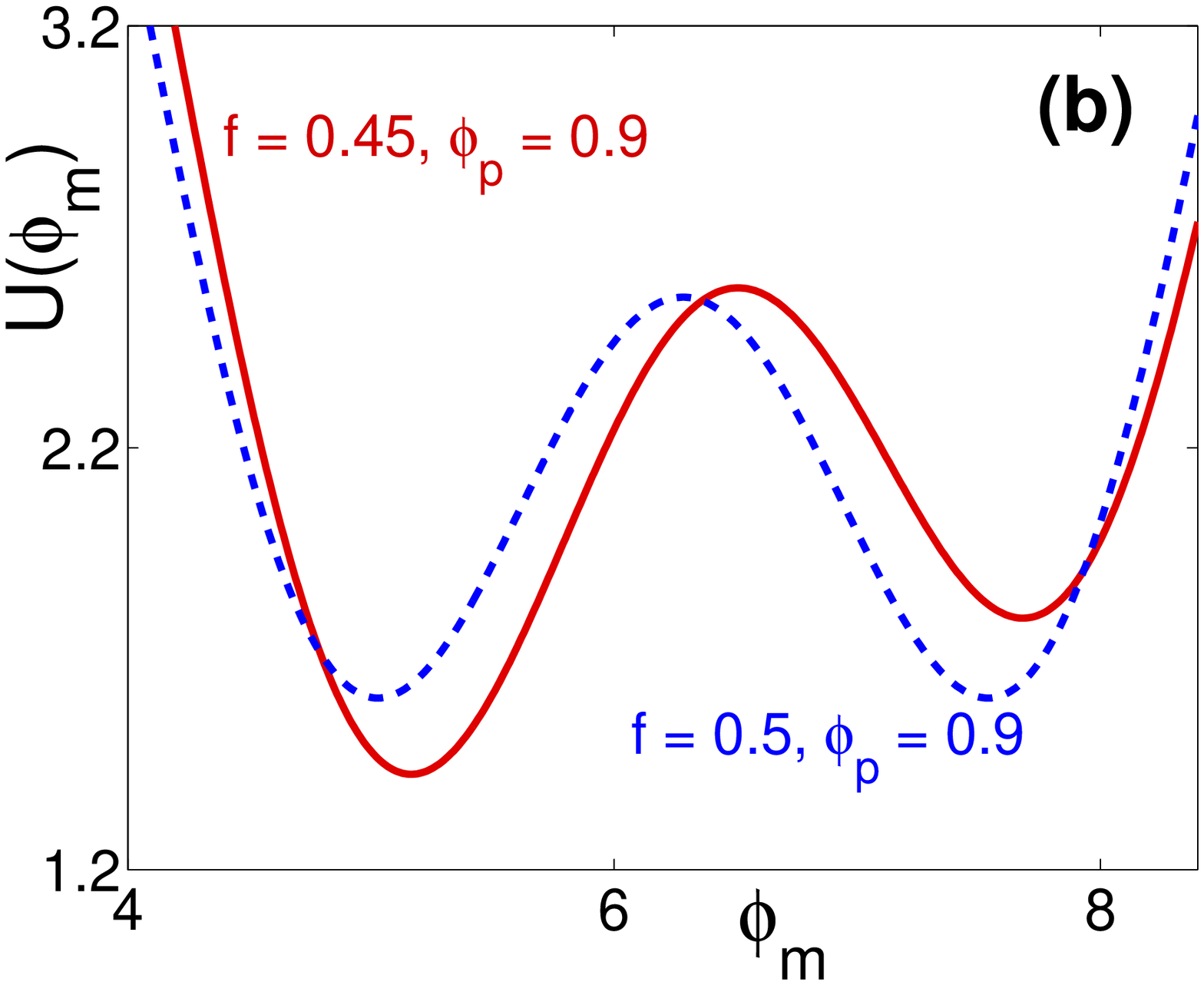}
\caption{(Color online) (a) A three-level artificial ``atom" made
of a superconducting loop, with three junctions, threaded by a
bias flux $\Phi_{e}$ and external field $\Phi_{a}$, consisting of
the quantized and time-dependent magnetic fluxes. (b) Potential
energy $U(\varphi_{p},\varphi_{m})$ versus the phase $\varphi_{m}$
for fixed $\varphi_{p}$ (e.g., $\varphi_{p}=0.9$) and reduced
magnetic flux $f=\Phi_{e}/\Phi_{0}=0.5$, for the dashed blue
curve, and $f=0.45$, for the continuous orange curve.}
\label{fig1}
\end{figure}

The artificial atom~\cite{you} considered here, described in
Fig.~\ref{fig1}(a), is a superconducting loop with three Josephson
junctions~\cite{orlando,yu,saito}. Two junctions have the same
Josephson energies and capacitances, which are $\alpha$ times
larger than that of the third one. Then, the Hamiltonian can be
written as~\cite{liu,orlando}
\begin{equation}
H^{\prime}\,\,=\,\,\frac{P_{p}^{2}}{2M_{p}}+\frac{P_{m}^{2}}{2M_{m}}
+U(\varphi_{p},\varphi _{m}, f),
\end{equation}
with the effective masses $M_{p}=2C_{\mathrm{J}}(\Phi_{0}/2\pi)^2$
and $M_{m}=M_{p}(1+2\alpha)$. The effective potential $U(\varphi
_{p},\varphi _{m},f)$ is
\begin{eqnarray}
U(\varphi _{p},\,\varphi _{m},\, f)&=& 2E_{\mathrm{J}}(1-\cos
\varphi _{p}\cos
\varphi _{m})  \notag \\
&+&\alpha E_{\mathrm{J}}[1-\cos (2\pi f+2\varphi _{m})]
\end{eqnarray}
where $\varphi _{p}=(\varphi _{1}+\varphi _{2})/2$ and $\varphi
_{m}=(\varphi _{1}-\varphi _{2})/2$ are defined by the phase drops
$\varphi_{1}$ and $\varphi _{2}$ across the two larger junctions;
$f=\Phi _{\mathrm{e}}/\Phi _{0}$ is the reduced bias magnetic flux
through the qubit loop, and $\Phi _{0}$ is the magnetic flux
quantum.

The potential energy $U(\phi _{p},\phi _{m},f)$ is an even
function of the canonical variable $\phi_{p}$, and naturally has
the mirror symmetry for $\phi_{p}\rightarrow=-\phi_{p}$. For other
variable $\phi _{m}$, the symmetry is completely determined by the
reduced bias magnetic flux $f$. This is shown in
Fig.~\ref{fig1}(b), comparing $f=0.5$ and $f=0.45$, for a given
$\phi _{p}=0.9$. When $2f=n$ with an integer $n$, the potential
energy $U$ has an inversion symmetry with respect to both phase
variables $\phi _{m}$ and $\phi _{p}$; that is,
\begin{equation}
U(-\phi _{m},\,-\phi _{p},\, 2f=n)\,=\,U(\phi _{m},\,\phi
_{p},\,2f=n),
\end{equation}
and thus the parities of the eigenstates are well-defined.
However, the inversion symmetry with $\varphi _{p}$ and $\varphi
_{m}$ is broken when $2f \neq n$,  that is,
\begin{equation}
U(-\phi _{m},\,-\phi _{p},\,2f \neq n)\,\neq \,U(\phi _{m},\,\phi
_{p},\,2f=n).
\end{equation}
Ref.~\cite{liu} computed the $f$-dependent energy spectrum, with
the lowest three energy levels, denoted by $|b\rangle $,
$|c\rangle $, and $|e\rangle $, well separated from the other
upper-energy levels. Since microwave-assisted transitions can
occur among the lowest three energy levels~\cite{liu}, this
artificial atom allows, cyclic or $\Delta $-shaped, transitions
when $f\neq 0.5$.

Besides the bias magnetic flux $\Phi_{e}$, we also apply another
magnetic flux $\Phi_{a}$, consisting of a quantized field and two
classical fields. To realize the strong coupling of the flux qubit
to a quantized field, now the flux qubit is coupled to a
one-dimensional transmission line resonator. This can be realized
by replacing the charge-qubit in the circuit QED
architecture~\cite{liu-epl,wallraff,liu-pra,yliu} by a flux qubit.
Then a single-mode quantized magnetic field can be provided by the
transmission line resonator.  All three fields are assumed to
induce transitions among the lowest three energy levels of the
artificial atom to form the $\Delta$-shaped configuration
mentioned above. The frequencies of the quantized and two
classical fields are assumed to be $\,\omega$, $\Omega_{e}$, and
$\,\Omega _{c}$, respectively.

\begin{figure}
\includegraphics[bb=44 300 555 730, width=5cm,clip]{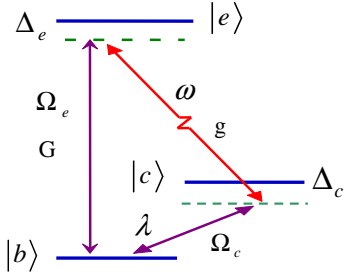}
\caption{(Color online).  Triangle- or $\Delta$-shaped transitions
among three energy levels of the ``artificial atom". The classical
fields with frequencies $\Omega_{e}$ and $\Omega_{c}$ induce the
transitions $(G,\,\lambda)$ with Rabi frequencies $G$ and
$\lambda$, while the quantized field with frequency $\omega$
induces the transition with Rabi frequency $g$.} \label{fig2}
\end{figure}

The Hamiltonian of the three-level artificial atom interacting
with the three fields can be written as
\begin{eqnarray}\label{eq:2}
H &=&\omega _{e}|e\rangle \langle e|+\omega _{c}|c\rangle \langle
c|+\omega a^{\dagger }a   \\
&+&(g|e\rangle \langle c|a+Ge^{i\Omega _{e}t}|b\rangle \langle e|
+\lambda e^{i\Omega _{c}t}|b\rangle \langle c|+{\rm
H.c.}).\nonumber
\end{eqnarray}
Here, we take $\hbar =1$. The quantized field is assumed to couple
the transition between $|e\rangle $ and $|c\rangle $, while the
two classical fields are applied between $|e\rangle $ and
$|b\rangle $, as well as between $|c\rangle $ and $|b\rangle $,
respectively.  $\omega _{e}$ ($\omega _{c}$) are transition
frequencies between $|e\rangle $ ($|c\rangle $) and $|b\rangle $
(see the Fig.~\ref{fig2}).  The detuning between the transition
frequency $\omega_{e}$ (or $\omega_{c}$) and the frequency of the
classical field $\Omega_{e}$ (or $\Omega_{c}$) is denoted by
\begin{equation}
\Delta _{e}=\omega_{e}-\Omega_{e} \,\,\,\,{\rm or} \,\,\,\,\Delta
_{c}=\omega _{c}-\Omega _{c}.
\end{equation}
$a$ and $a^{\dagger }$ are the annihilation and creation operators
of the quantized mode, $G$ and $\lambda $ are the Rabi-frequencies
of the classical fields, $g$ denotes the vacuum Rabi-frequency of
the quantized mode. Without loss of generality, we assume that all
Rabi frequencies are real numbers. Here, we assume that the
frequencies of the three fields satisfy the condition
\begin{equation}
\Omega _{e}-\Omega _{c}=\omega.
\end{equation}
This condition is required such that the equivalent Hamiltonian in
a ``rotating" reference frame (defined below) will be
time-independent. In this case, the evolution of the quantum
system will remain in the adiabatic subspace when the Rabi
frequencies are adiabatically changed to transfer the quantum
information, carried by photons, to the artificial atoms.

Figure~\ref{fig2} illustrates the transitions induced by the
interactions of the artificial atom with the three fields. This
cyclic or $\Delta $-shaped transitions define a new type of atom,
different from the $\Lambda $ (or  $\Xi $, or $V$)-type
atoms~\cite{scullybook,jpm}. In a ``rotating" reference frame of a
time-dependent unitary transformation
\begin{equation}
W(t)=\exp [-it(\Omega _{e}|e\rangle \langle e|+\Omega_{c}|c\rangle
\langle c|+\omega a^{\dag }a)],
\end{equation}
the Hamiltonian in Eq.~(\ref{eq:2}) can be rewritten as
\begin{eqnarray}
H &=&\Delta _{c}\,|c\rangle \langle c|+\Delta _{e}\,|e\rangle
\langle
e|  \label{eq:4} \nonumber \\
&+&(g\,|e\rangle \langle c|a+G\,|e\rangle \langle b|+\lambda\,
|b\rangle \langle c|+{\rm H.c.}),
\end{eqnarray}
where the the frequencies-matching condition
$\Omega_{e}-\Omega_{c}=\omega$ has been used.

The population of the three-level artificial atom can be cyclically
transferred by adiabatically applying three classical fields~\cite{liu}.
However, in the presence of a quantized field, the transitions
\begin{eqnarray*}
|e,n\rangle &\leftrightarrow &|c,n+1\rangle \leftrightarrow
|b,n+1\rangle \\
&\leftrightarrow &|e,n+1\rangle \leftrightarrow |c,n+2\rangle
\leftrightarrow |e,n+2\rangle \,\cdots
\end{eqnarray*}
cannot form a closed cycle because each cycle produces a one
photon excitation. The triangular or $\Delta $-shaped geometry of
the transitions is shown in Fig.~\ref{fig3}, where the classical
fields can only induce transitions in the plane of each triangle
of atom-photon joint states, while the quantized field drives the
transitions from one plane to another, by increasing or decreasing
one photon.

\section{Mechanism to generate nonclassical photon states}

In this section, we will consider the possibility to utilize the
above $\Delta$-shaped three level artificial atom as a basic
single photon device. It is well known that there has been
considerable interest in the generation of non-classical light
using solid-state devices for highly sensitive metrology and
quantum information. Some solid-state lasers have been proposed to
emit non-classical light with photon number squeezing, but the
present proposal, based on $\Delta$-shaped artificial atoms, is
essentially a macroscopic quantum device,  which, in principle,
could be easily controlled by only using classical parameters
(e.g., the magnetic flux).

\begin{figure}
\includegraphics[bb=61 311 534 700, width=6 cm,clip]{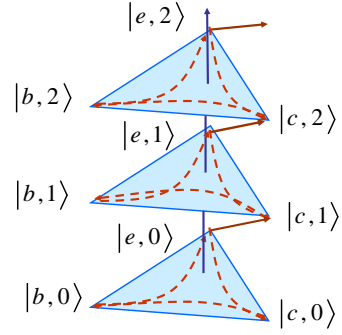}
\caption{(Color online). Each triangle has a different number
 (e.g., 1, 2, 3) of photons. The classical fields can only induce
in-plane transitions to form CT. The quantum field links different
triangle planes to generate photons. } \label{fig3}
\end{figure}

\begin{figure}
\includegraphics[bb=4 300 570 670, width=9 cm, clip]{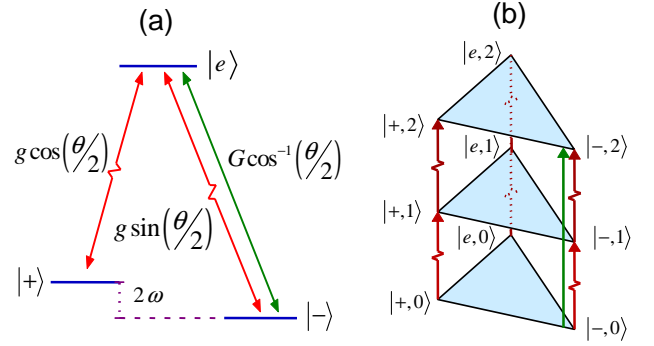}
\caption{(Color online) Three-level artificial atom in the
classical-field dressed picture. (a) A $\Lambda$-like atom
equivalent to the $\Delta$-atom in Fig.~\ref{fig2} with the
$|+\rangle\leftrightarrow|e\rangle$ transition coupled, via the
left zig-zag-line, to a displaced quantized field, denoted by the
operator $A$ in Eq.~(\ref{eq:8b}). The
$|-\rangle\leftrightarrow|e\rangle$ transition couples both an
equivalent classical field, denoted by $\eta$, and a displaced
quantized field $A$ in Eq.~(\ref{eq:8b}). (b) After doing
adiabatic elimination for large detuning, there are no transitions
among the three energy levels in the same equal-number-of-photons
triangle-plane. The vertical arrowed lines linking the vertices of
the triangles represent the transitions that accompany the
creation of photons.} \label{fig4}
\end{figure}

To intuitively describe the main mechanism of how to create the
quasi-classical and non-classical photon states by using the
transition configuration shown in Fig.~\ref{fig3}, we first
rewrite the sub-Hamiltonian in Eq.~(\ref{eq:4})
\begin{equation}\label{eq:10}
H_{s}=\Delta _{c}\,|c\rangle \langle c|+[\lambda |c\rangle \langle
b|+{\rm H.c.}],
\end{equation}
into
\begin{equation}
H_{s}=\epsilon_{+}\,|+\rangle \langle +|+\epsilon _{-}\,|-\rangle
\langle -|
\end{equation}
with two dressed states
\begin{eqnarray*}
\left\vert +\right\rangle &=&\cos \left(\frac{\theta
}{2}\right)\left\vert c\right\rangle+\sin \left(\frac{\theta
}{2}\right)\left\vert b\right\rangle,  \\
\left\vert -\right\rangle &=&-\sin \left(\frac{\theta
}{2}\right)\left\vert c\right\rangle +\cos \left(\frac{\theta
}{2}\right)\left\vert b\right\rangle,
\end{eqnarray*}
where we have defined the mixing angle
\begin{equation}
\theta =\arctan \left(\frac{2\lambda}{\Delta_{c}}\right).
\end{equation}
It is obvious that $\theta$ can be controlled through the detuning
$\Delta_{c}$ by changing the frequency of the classical field. The
states $\left\vert \pm \right\rangle $ are the eigenstates of
$H_{s}$ corresponding to the eigenvalues
\begin{equation}
\epsilon _{\pm }=\frac{\Delta _{c}}{2}\pm \omega ^{\prime },
\end{equation}
with the dressed frequency
\begin{equation}
\omega ^{\prime }=\sqrt{\lambda ^{2}+\frac{\Delta _{c}^{2}}{4}}.
\end{equation}
Then, in this dressed basis, the total Hamiltonian in
Eq.~(\ref{eq:4})
\begin{subequations}
\begin{equation}\label{eq:3}
H=H_{0}+H_{1}
\end{equation}
can be rewritten as
\begin{equation}
H_{0}=\Delta _{e}\,|e\rangle \langle e|+\epsilon _{+}\,|+\rangle
\langle +|+\epsilon _{-}\,|-\rangle \langle -|\,  \label{eq:8a}
\end{equation}
and
\begin{equation}
H_{1}=g(\theta )A\,|e\rangle \langle +|-G(\theta )B\,|e\rangle
\langle -|+{\rm H.c.} \label{eq:8b}
\end{equation}
with the displaced boson operators $A=a+\xi $ and $B=A-\zeta$, and
the controllable parameters
\end{subequations}
\begin{eqnarray*}
g(\theta ) =g\cos \left(\frac{\theta}{2}\right),&&
G(\theta ) =g\sin \left(\frac{\theta}{2}\right), \\
\xi (\theta) =\frac{G}{g}\tan\left(\frac{\theta}{2}\right),&&
\zeta(\theta)=\frac{G}{g}\tan^{-1}\left(\frac{\theta}{2}\right).
\end{eqnarray*}
The Hamiltonian~(\ref{eq:3}) describes the $\Lambda $-like
transition atom shown in Fig.~\ref{fig4}(a). Instead of the usual
$\Lambda $-type atom, the transitions between states $|e\rangle $
and $|-\rangle $ are induced by two fields, one is a quantized
light field with coupling strength $g\sin (\theta/2)$, described
by a displaced annihilation operator $a$, another is a classical
field with the Rabi frequency $G\cos(\theta/2)$.

Figure~\ref{fig4}(a) schematically describes the creation of
quasi-classical and non-classical photon states based on the CT
process. Due to the coherent $|c\rangle $-$|b\rangle $ interaction
with the coupling strength $\lambda$, as in Eq.~(\ref{eq:10}), the
system can be described by the driven JC model shown in
Eq.~(\ref{eq:3}). However, for large detunings, in
Eqs.~(\ref{eq:3}-\ref{eq:8b}),  i.e., $\Delta
_{e}-\epsilon_{\pm}\gg g(\theta), \,G(\theta)$, we can
adiabatically separate the excited state $|e\rangle $ and then a
coherent transition between states $|c\rangle $ and $|b\rangle $
is induced by the quantized field originally applied between
$|e\rangle $ and $|c\rangle $. This is very similar to the usual
Jaynes-Cummings (JC) model, obtained by adiabatically eliminating
the highest third energy level in the stimulated Raman scattering
of intense laser light. In the dressed states $|\pm \rangle $
basis, and for the above large detuning condition, there exist
three types of subspaces, related to states $|+\rangle $,
$|-\rangle $, and $|e\rangle $, respectively. These subspaces are
depicted in Fig.~\ref{fig4}(b) by vertical lines linking the
vertices of the triangles. Corresponding to each state, e.g.,
$|+\rangle $, the photon mode is driven by an effective external
force depending on the coherent $|c\rangle $-$|b\rangle $
interaction, and thus the single mode photon states can be
produced from the vacuum state.

\section{Adiabatic generation of Schr\"{o}dinger cat states}

In order to better understand the above-mentioned mechanism to
generate non-classical photon states from these controllable
artificial atoms, we demonstrate the adiabatic generation of
Schr\"odinger cat states. In the large detuning limit, we can
adiabatically eliminate the terms causing transitions from
$|e\rangle $ to $|+\rangle $ and $|-\rangle$.

The adiabatic elimination can be done by using the
Fr\"ohlich-Nakajima transformation (FNT)~\cite{fro,nakajima},
which is applied to achieve the effective electron-electron
interaction Hamiltonian in the BCS theory. To consider the
validity of this method, we will show that it is equivalent to a
result of the second order perturbation in the Appendix A. In the
FNT method, we define a transformation by the operator $V=$ $\exp
(S)$, with an anti-Hermitian operator $S$ to be determined. Then
we apply this transformation $V$ to the original
Hamiltonian~(\ref{eq:3}) to given an equivalent Hamiltonian
$H_{V}=V^{\dagger }HV.$

We assume that the operator $S$ to be the perturbation term with
the same order as $H_{I}$, and then we can expand $H_{V}$ in the
series of $S$. In general, we can consider the Hamiltonian of an
interacting system, described by a sum of free Hamiltonian $H_{0}$
and the interaction Hamiltonian $H_{1}$ as $H=H_{0}+H_{1}$, shown
in Eq.~(\ref{eq:3}). By comparing with the free part $H_{0}$, the
interaction part $H_{1}$ can be regard as a perturbation term. Let
us perform the transformation $V=$ $\exp (S)$ on the Hamiltonian
$H=H_{0}+H_{1}$. Then, we can derive an approximately equivalent
Hamiltonian $H_{V}$ as
\begin{equation}\label{eq:13}
H_{V}\cong H_{0}+\frac{1}{2}\left[H_{1},\,S\right],
\end{equation}
where the operator $S$ can be determined by
\begin{equation}\label{eq:h}
H_{1}+[H_{\mathrm{0}},S]=0.
\end{equation}

The transformation, by which one can obtain the effective
Hamiltonian in Eq.~(\ref{eq:13}) from the Hamiltonian in
Eq.~(\ref{eq:3}),  is the so-called generalized Fr\"ohlich
transformation (for details, see Appendix A).

If we replace $H_{0}$ and $H_{1}$ in Eq.~(\ref{eq:h}) by
the explicit expressions in Eqs.~(\ref{eq:8a}) and (\ref{eq:8b}),
and assume
\begin{eqnarray}
S &=&\Gamma _{1}\,A\,|e\rangle \langle +|+\Gamma
_{2}\,B\,|e\rangle
\langle -| \notag  \label{eq:12} \\
&+&\Gamma _{3}\,A^{\dagger}\,|+\rangle \langle e|+\Gamma
_{4}\,B^{\dagger }\,|-\rangle \langle e|,
\end{eqnarray}
for parameters $\Gamma _{i}\,(i=1,\,2,\,3,\,4)$ to be determined,
then the parameters $\Gamma _{i}\,(i=1,\,2,\,3,\,4)$ can be
obtained as
\begin{subequations}
\begin{eqnarray}
\Gamma _{1} &=&-\Gamma _{3}\,=\,-\frac{g(\theta )}{\epsilon +\Delta },\\
\Gamma _{2} &=&-\Gamma _{4}\,=\,-\frac{G(\theta )}{\epsilon
-\Delta },
\end{eqnarray}
\end{subequations}
with
\begin{equation}
\Delta=\frac{1}{2}\sqrt{(\Delta_{c}-\Delta_{e})^2+4\lambda^2},\,\,\,
\epsilon=\frac{1}{2}(\Delta_{c}+\Delta_{e}).
\end{equation}
Then, using the expressions of $S$, $H_{0}$, and $H_{1}$ in
Eqs.~(\ref{eq:12}), (\ref{eq:8a}), and (\ref{eq:8b}), we can
obtain an effective Hamiltonian from Eq.~(\ref{eq:13}) as
\begin{subequations}
\begin{equation}
H_{V}\approx H_{e}|e\rangle \langle e|+H_{bc}.
\end{equation}
Here, the Hamiltonians $H_{e}$ and $H_{bc}$ can be expressed as
\begin{equation}
H_{e}=\Delta _{e}+\Omega _{A}\,A\,A^{+}+\Omega _{B}\,B\,B^{+}
\end{equation}
and
\begin{eqnarray}
H_{bc} &=&(\epsilon _{+}-\Omega _{A}A^{+}A)|+\rangle \langle +|  \notag \\
&&+(\epsilon _{-}-\Omega _{B}B^{+}B)|-\rangle \langle -|  \notag \\
&&+\Gamma \lbrack AB^{+}|+\rangle \langle -|+A^{+}B|-\rangle \langle +|],
\end{eqnarray}
with
\end{subequations}
\begin{equation}
\Gamma =\frac{G(\theta )g(\theta )}{2\Delta _{-}\Delta
_{+}}(2\Delta _{e}-\Delta _{c}).
\end{equation}
The effective frequencies
\begin{equation}\label{eq:21}
\Omega _{A}=\frac{g^{2}(\theta )}{\Delta _{+}},\,\,\,\,\Omega
_{B}=\frac{ G^{2}(\theta )}{\Delta _{-}}
\end{equation}
represent the Stark shifts with $\Delta _{\pm }=\Delta
_{e}-\epsilon _{\pm }$.

According to former definitions of the operators $A$ and $B$ in
Eq.~(\ref{eq:8b}), the Hamiltonian $H_{e}$ can be rewritten as
\begin{eqnarray}
H_{e} &=&(\Omega _{A}+\Omega _{B})\,a\,a^{\dagger } \\
&&+[(\xi \,\Omega _{A}-\eta \,\Omega _{B})\,a^{\dagger }+{\rm
H.c.}] \notag
\end{eqnarray}
after neglecting the constant terms $\Delta _{e}+|\xi |^{2}+|\eta
|^{2}$. It is clear that the Hamiltonian $H_{e}$ describes a
driven harmonic oscillator. Then, when the total system can be
adiabatically kept in the excited state $|e\rangle $, $H_{e}$
describes the creation of a coherent photon state from the
vacuum~\cite{scullybook}. However, due to the spontaneous emission
of excited states, it is difficult to keep the artificial atom in
its excited state $|e\rangle $. Thus, let us now consider how to
generate non-classical photon states by only using the more robust
lower states $|\pm \rangle $.

The last term in $H_{bc}$ oscillates in a larger frequency range:
$|\epsilon_{+}-\epsilon _{-}|\simeq 2\omega ^{\prime }$. Thus, in
the rotating wave approximation, we have
\begin{eqnarray}
H_{bc}&=&(\epsilon _{+}-\Omega _{A}A^{+}A)|+\rangle
\langle+|\nonumber\\
&+&(\epsilon _{-}-\Omega _{B}B^{+}B)|-\rangle \langle -|.
\end{eqnarray}
This is the standard Hamiltonian to describe the dynamical
generation of Schr\"odinger cat states (e.g., Ref.~\cite{cqed}).
Since the bare ground state $|b\rangle $ is easy to be
initialized, we can assume that the artificial atom is initially
in the bare ground state $|b\rangle =\sin (\theta /2)|+\rangle
+\cos (\theta /2)|-\rangle $, while the cavity field is initially
in the vacuum state $|0\rangle $. Then at time $\tau$, the whole
system can evolve into
\begin{eqnarray}
|\psi (\tau )\rangle \,\, &=&\exp (iH_{bc}t)[\sin (\theta
/2)|+\rangle +\cos
(\theta /2)|-\rangle ]|0\rangle  \notag \\
&=&\sin \frac{\theta }{2}\exp [i\xi ^{2}\exp (-i\Omega _{A}t)]
|\alpha (-\xi,t)\rangle |+\rangle  \notag \\
&+&\cos \frac{\theta
}{2}\exp [i\zeta ^{2}\exp (-i\Omega _{B}t)]|\alpha (\zeta
,t)\rangle |-\rangle .
\end{eqnarray}
where $|\alpha (x,t)\rangle =$ $|\alpha =\alpha (x,t)\rangle $
(and $x=\xi,\zeta $) denotes coherent states with
\begin{equation}
\alpha (x,t)=x[1-\exp (i\Omega _{x}t)]
\end{equation}
and $\Omega _{\xi }=\Omega _{A,}$ $\Omega _{\zeta }=\Omega _{B}$.
By adjusting the coupling constant $\lambda $ between $|c\rangle $
and $|b\rangle $, in this ``cyclic atom", one can control
dynamical processes to obtain the cat states of the qubit
subsystem consisting of the two dressed states $|\pm \rangle $
entangled with the quantized field.

To show the existence of the ``cat", we need to calculate the
overlap
\begin{equation}
F(\lambda ,t)=|\langle \alpha (\zeta ,t)|\alpha (-\xi ,t)\rangle |=\exp
[-y(\lambda ,t)]  \label{eq:7}
\end{equation}
for two coherent states $|\alpha (-\xi ,t)\rangle $ and $|\alpha
(\zeta ,t)\rangle $, where
\begin{eqnarray}
y(\lambda ,t) &=&2(\zeta +\xi )^{2}-4\zeta \xi \sin ^{2}\left[ \frac{t}{2}%
(\Omega _{B}-\Omega _{A})\right]  \label{eq:8} \\
&&-2\zeta (\zeta +\xi )\cos (\Omega _{B}t)-2\xi (\zeta +\xi )\cos (\Omega
_{A}t).  \notag
\end{eqnarray}

\begin{figure}[tbp]
\includegraphics[bb=37 188 563 599, width=7.5cm,clip]{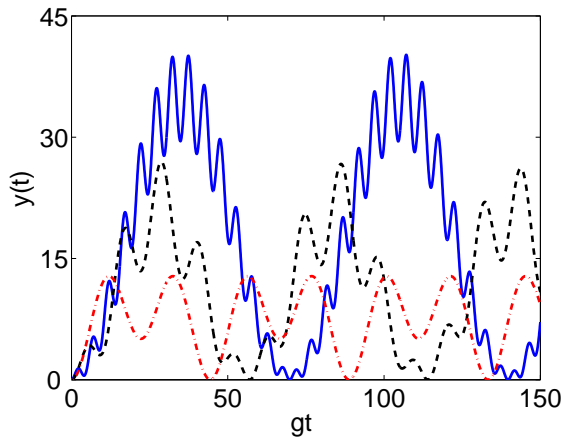}
\caption{(Color online). The exponent $y(t)$, see
Eqs.~(\ref{eq:7}) and ~(\ref{eq:8}), of the overlap of two
coherent states is plotted as a function of $gt$, see
Eqs.~(\ref{eq:21}), ~(\ref{eq:7}) and ~(\ref{eq:8}) for given
parameters $\Delta_{e}=3\lambda$, $G=0.9\lambda$, $g=0.8\lambda$,
$\theta=\protect\pi/2$ (dash-dotted red curve), $\theta=\pi/3$
(dotted black curve), and $\theta=\pi/4$ (solid blue curve).}
\label{fig5}
\end{figure}

In Fig.~\ref{fig5}, the time evolution of $y(t)$ is plotted for
given parameters, e.g., $\Delta _{e}=3\lambda $, $G=0.9\lambda $,
$g=0.8\lambda $ for different values of $\theta =\arctan (2\lambda
/\Delta _{c})=\pi /2,\,\,\pi /4,\,\,\pi /6$. It shows that $y(t)$
can periodically reach its maximum value, which means that
$|\langle \alpha (\zeta ,t)|\alpha (-\xi ,t)\rangle |$ becomes
minimum at these times, with period $2\pi $. The period of the
function $y(t)$ is determined by three frequencies $\Omega _{A}$,
$\Omega _{B}$, and $\Omega _{B}-\Omega _{A}$, so Fig.~\ref{fig5}
shows the small modulation overimposed on the larger modulation.
We find that a larger detuning $\Delta _{c}$ corresponds to a
larger maximum value when other parameters are fixed. However,
$y(t)$ needs a longer period to reach these maximum points. The
above result demonstrates that macroscopic Schr\"odinger cat
states, an entanglement between a \textit{macroscopic} quantum
two-level system (macroscopic qubit) and the non-classical photon
states, can be generated by superconducting quantum devices. These
cat states are \textit{different} from the usual Schr\"odinger cat
states, an entanglement between a \textit{microscopic} two level
atom and the quasi-classical photon states, which are created by
using the atomic cavity QED~\cite{cqed}.

The above setup can also be used to create a coherent state if the
cyclic artificial atom is initially prepared in the state
$|-\rangle $. In this case, the total system should be
adiabatically kept in the ground state $|-\rangle $. The effective
Hamiltonian becomes $H_{-}\simeq -\Omega _{B}B^{+}B$, by dropping
the constant term. More explicitly,
\begin{equation}
H_{-}\simeq -\Omega _{B}a^{+}a+f(\theta )(a+a^{+})  \label{eq:20}
\end{equation}
realizes a driven harmonic oscillator.  The driving force
$f(\theta )$ can be expressed as $f(\theta )=G(\theta
)g(\theta)/\Delta _{-}$, and it depends on the coupling constant
$\lambda$. Starting from the vacuum $|0\rangle$, with a duration
$t$, the single-mode quantized field will evolve into a coherent
state $|\varphi (t)\rangle =|\alpha \rangle $ with $\alpha =\zeta
\lbrack 1-\exp (i\Omega _{B}t]$, where a time-dependent global
phase $\exp [i\zeta ^{2}(\sin (\Omega _{B}t)-\Omega _{B}t)]$ has
been neglected. From the expression of the photon number
\begin{equation}
N(t)\,=\,\langle\alpha| a^{+}a|\alpha\rangle\, =\,\zeta
^{2}|1-\exp (i\Omega _{B}t))|^{2},
\end{equation}
we can calculate the generation rate of the photons in the
quantized mode:
\begin{equation}
r(t)\,=\,\left\vert \frac{\mathrm{d}N(t)}{\mathrm{d}t}\right\vert
=\frac{2g^{2}(\theta )}{\Delta _{-}}\sin (\Omega _{b}t).
\end{equation}
This result shows that, the coupling strength $\lambda $ of the
interaction between $\left\vert c\right\rangle $ and $\left\vert
b\right\rangle $, caused by the symmetry-breaking, can be used to
enhance the probability of creating single-mode photons. If there
is no interaction between $|c\rangle $ and $|b\rangle $, the
external force $f(\theta )$ would vanish accordingly and then the
dynamic evolution cannot automatically produce coherent photon
states.

\section{Conclusions}

In an artificial atom represented by a superconducting quantum
circuit, we briefly review optical transitions and their selection
rules. It is shown that all transitions are possible in such
artificial atom when the applied bias magnetic flux is not at the
optimal point~\cite{liu}. Then cyclic or $\Delta$-shaped
transitions can be realized for the lowest three energy levels in
this artificial atom. Using this cyclic population transfer
mechanisms, we have studied how to create nonclassical single-mode
photon states and a macroscopic Schr\"odinger cat states. We show
that this approach is controllable, because either the ground
state or the two lowest-energy levels are utilized through their
coherent coupling to external fields, which can be used to control
the parameters of the system. For example, if the Rabi frequency
$\lambda=0$,  then the classical field, which induces transitions
between states $|c\rangle$ and $|b\rangle$, is set to zero. Thus,
our model can be referred to  $\Lambda-$type atom driven by a
quantized and a classical fields~\cite{parkins}. In this
case~\cite{parkins}, neither the cat state nor the coherent state
can be generated from the initial state of the whole system with
the bare ground state of atom and the vacuum of the quantized
field. Starting from either the vacuum or a coherent state, it is
a deterministic scheme to generate nonclassical photon states via
CT manipulations.

The large detuning limit implies a relatively weak coupling
constants $g(\theta)$ and $G(\theta)$, as well as relatively large
detuning $\Delta_{e}-\epsilon_{\pm}$, shown in
Eqs.~(\ref{eq:8a}-\ref{eq:8b}). Thus it limits the efficient
adjustment of the dynamic processes. Therefore, a generic version
of our proposal might not be very efficient. Fortunately, in our
proposal,  the strength $\lambda$ of the controlling-field
coupling between $|c\rangle $ and $|b\rangle$ is adjustable. Thus,
one can feasibly manipulate this parameter for our goal without
violating the requirement of large detuning.

Another question is the problem of decoherence. An efficient
scheme requires the decoherence time comparable with the
characteristic time of the effective frequencies $\Omega _{B}$ and
$\Omega _{A}$. It is known that the interaction strength $g$
between the qubit and quantized field can reach about $100$ MHz if
we use the transmission line resonator in the circuit
QED~\cite{gir,wallraff}. According to the definitions of $\Omega
_{B}$ and $\Omega _{A}$ in Eq.~(\ref{eq:21}), they can be of the
order of $10-100$ MHz, if we choose appropriate detunings $\Delta
_{e}$ and $\Delta _{c}$. Then, it is possible to realize our
proposal within the experimental values~\cite{gir1} for $T_{1}\sim
7\,\mu $s and \,$T_{2}\sim 800$ ns.

Finally, we should point out the relation between our present work
and the quantum Carnot engine (QCE) in Ref.~\cite{scully,quan}. In
the QCE proposal, the $\Lambda $-type atoms are prepared as a
superposition of two lower states. In the $\Delta$-type transition
configuration, the superposition of the two lowest states can be
naturally produced by the interaction between the field and the
artificial atom, and hence the cyclic three-level atom is a good
candidate to demonstrate the QCE.

\section{Acknowledgments}

We acknowledge the partial support of the US NSA and ARDA under
AFOSR contract No. F49620-02-1-0334, and the NSF grant No.
EIA-0130383. The work of CPS is also partially supported by the
NSFC and Fundamental Research Program of China with No.
2001CB309310.

\appendix
\section{Generalized Fr\"ohlich-Nakajima transformation and its
equivalence to perturbation theory }

Let us consider a Hamiltonian $H$ of a given system with its free
part $H_{0}$ and a perturbation term $H_{I}$
\begin{equation}\label{eq:A1}
H=H_{0}+\lambda H_{I}.
\end{equation}
Here, $\lambda $ is the so-called perturbation parameter
introduced to characterize the order of the perturbation. At the
end of calculation, $\lambda$ is taken as unity.

The crucial point of the generalized Fr\"ohlich-Nakajima
transformation is to choose a proper unitary transformation
$V(\lambda)= \exp(\lambda S)$, where $S$ is an anti-Hermitian
operator, to be determined. The inverse transformation of
$V(\lambda)$ makes the states $|\Psi\rangle$, governed by the
Hamiltonian in Eq.~(\ref{eq:A1}), change to a new state
\begin{equation}
|\Phi\rangle=V(-\lambda)|\Psi\rangle=\exp(-\lambda S)|\Psi\rangle.
\end{equation}
And the evolution of the state $|\Phi\rangle$ is governed by the
transferred Hamiltonian
\begin{equation}
H_{\lambda}=e^{-\lambda S}He^{\lambda S}.
\end{equation}
It is well known that the unitary transformation does not change
the dynamics of the system, and then the Hamiltonians
$H_{\lambda}$ and $H$ describe the same physical process. Here the
operator $S$ should be appropriately chosen such that it has the
same order as the perturbation term $H_{I}$. Physically, the
effect of the Hamiltonian $H_{I}$ on the final result is so small
that it can be neglected.

Using the Baker-Campbell-Hausdorff formula, the Hamiltonian
$H_{\lambda }$ can be expressed in a series of the parameter
$\lambda$ as
\begin{equation}\label{eq:A4}
H_{\lambda }=H_{0}+\sum_{n=1}\frac{\lambda ^{n}(-1)
^{n-1}}{(n-1)!}{\underset{n-1}{\underbrace{[S,\, [ S,\,\cdots [
S}},\,\,H_{I}] ] ] }.
\end{equation}
Second order perturbation theory can be realized by imposing the
condition
\begin{equation}\label{eq:A5}
H_{I}+\left[ H_{0},S\right] =0
\end{equation}
on Eq.~(\ref{eq:A4}). Eq.~(\ref{eq:A5}) can be used to determine
the operator $S$. For the sake of simplicity, the eigenstates of
$H_{0}$ are assumed to be non-degenerate. Let $|n\rangle $ be the
eigenstate of the Hamiltonian $H_{0}$ with the eigenvalue $E_{n}$.
Taking the matrix elements of Eq.~(\ref{eq:A5} ) with respect to
the basis $\{ |n\rangle \}$ as
\begin{equation}
\langle m|H_{I}|n\rangle +\left( E_{m}-E_{n}\right) \langle
m|S|n\rangle =0,
\end{equation}
we can find the explicit expression of matrix elements for the
operator $S$
\begin{equation}
S_{mn}=\langle m|S|n\rangle =\frac{\langle m|H_{I}|
n\rangle}{E_{n}-E_{m}}.
\end{equation}
Thus, the representation of the operator $S$ in the  $\{ |n\rangle
\}$ basis can be
\begin{equation}
S=\sum_{m\neq n}\frac{\langle m|H_{I}| n\rangle}{E_{n}-E_{m}}\,
|m\rangle \langle n|.
\end{equation}

From Eqs.~(\ref{eq:A4}) and (\ref{eq:A5}), we obtain the effective
Hamiltonian
\begin{equation}\label{eq:A9}
H_{\lambda}\cong H_{0}+\frac{1}{2}\left[ H_{I}, \,S\right]
\end{equation}
up to second order in $H_{I}$. Using a matrix representation,
$H_{\lambda}$ can be expressed as
\begin{eqnarray}\label{eq:A10}
H_{\lambda}&=&\sum_{n} E_{n}|n\rangle \langle
n|\nonumber\\
&+&\sum_{l\, (l\neq n),\,m }\frac{\langle m|H_{I}|
l\rangle\,\langle l|H_{I}| n\rangle}{ 2(E_{n}-E_{l})}\,|m\rangle
\langle n|
\end{eqnarray}
in the $\{|n\rangle \}$ basis. We can see that the
Fr\"ohlich-Nakajima transformation is only applicable to a systems
with $\langle m|H_{I}|m\rangle=0$. Actually we can decompose the
total Hamiltonian $H$ such that $H_{0}$ only includes all diagonal
elements in the $\{|n\rangle \}$  basis of eigenstates for the
Hamiltonian $H_{0}$ while the off-diagonal ones are included in
$H_{I}$.

It is easy to obtain the eigenvalues of the transferred
Hamiltonian in Eq.~(\ref{eq:A9}) or (\ref{eq:A10}), up to second
order in $H_{I}$,  as
\begin{eqnarray}\label{eq:A11}
E_{n}^{(0)} &=&\langle n|H_{0}|n\rangle
+\frac{1}{2}\langle n|\left[ H_{I},\,\,S\right] |n\rangle   \notag \\
&=&E_{n}+\sum_{l\neq n}\frac{\vert \langle l| H_{I}|
n\rangle\vert^{2}}{E_{n}-E_{l}},
\end{eqnarray}
which correspond to the zero-order eigenstates of the Hamiltonian
$H_{\lambda}$. The second term in the right side of
Eq.~(\ref{eq:A11}) is the so-called self-energy term.

In fact,  from Eq.~(\ref{eq:A10}), it can be found that zero-order
eigenstates $|\Psi_{n}^{(0)}\rangle$ of the Hamiltonian
$H_{\lambda}$ are just the eigenstates $|n\rangle$ of the
Hamiltonian $H_{0}$, i.e., $|\Psi_{n}^{(0)}\rangle=|n\rangle$. The
eigenvalues in Eq.~(\ref{eq:A11}) provide energy corrections using
the time-independent perturbation theory.

To consider the relation between the Fr\"ohlich-Nakajima
transformation and the time-independent perturbation theory, we
can transfer eigenstates $|\Psi_{n}^{(0)}\rangle$ back to the
original picture. In this case, the first order eigenstates $|\Psi
_{n}^{(1)}\rangle$ of the Hamiltonian $H$ can be obtained by
\begin{eqnarray}
|\Psi _{n}^{(1)}\rangle &=& V\left( \lambda \right)
|\Psi_{n}^{(0)}\rangle
=\left( 1+S\right) |n\rangle  \\
&=&|n\rangle +\sum_{m\neq n}\frac{\langle m|H_{I}|n\rangle }{
E_{n}-E_{m}}|m\rangle,   \notag
\end{eqnarray}
where the expansion $V\left( \lambda \right)$ is kept up to first
order in $\lambda$.

It is easy to prove that $|\Psi_{n}^{(1)}\rangle$ are just the
first-order eigenstates of the original Hamiltonian $H$, with
respect to the perturbation decomposition of $H_{0}$ and $H_{I}$.
Since we have chosen that $H_{I}$ does not have diagonal terms,
the first correction to the energy is zero, and then $E_{n}$ is
also the result of the first correction of the energy for the
Hamiltonian $H$.

The eigenvalues in Eq.~(\ref{eq:A11}) are up to the second order
corrections. Correspondingly, the eigenstates $|\Psi _{n}^{ (2)
}\rangle$ of $H$ corresponding to the second-order energy
corrections can be given by acting $V(\lambda)$ on the first order
eigenstates $|\Psi_{n}^{(1)}\rangle$ of the Hamiltonian
$H_{\lambda}$. That is

\begin{eqnarray}
|\Psi _{n}^{ (2) }\rangle  &=&V(\lambda )\, |\Phi_{n}^{(1)}\rangle
=V^{-1}(\lambda )\, |\Psi_{n}^{(1)}\rangle\nonumber\\
&=&\left( 1+S+\frac{S^{2}}{2}\right)
|\Psi _{n}^{(1)}\rangle  \notag \\
&=&|n\rangle +\sum_{m\neq n}\frac{\langle m|
H_{I}|n\rangle}{E_{n}-E_{m}}|m\rangle  \\
&+&\sum_{l,\,m,\,l\neq n}\frac{\langle m| H_{I}|l\rangle\langle l|
H_{I}|n\rangle}{2\left( E_{l}-E_{m}\right) \left(
E_{n}-E_{l}\right)}|m\rangle.   \notag
\end{eqnarray}

\end{document}